\documentclass[preprint,nofootinbib]{revtex4}%
\usepackage{amssymb}
\usepackage{amsfonts}
\usepackage{amsmath}
\usepackage{graphicx}%
\providecommand{\U}[1]{\protect\rule{.1in}{.1in}}
\begin{document}
\title{Higher spin gravity in 3D:\\black holes, global charges and thermodynamics}
\author{Alfredo P\'{e}rez$^{1}$, David Tempo$^{1}$, Ricardo Troncoso$^{1,2}$}
\email{aperez, tempo, troncoso@cecs.cl}
\affiliation{$^{1}$Centro de Estudios Cient\'{\i}ficos (CECs), Casilla 1469, Valdivia, Chile}
\affiliation{$^{2}$Universidad Andr\'{e}s Bello, Av. Rep\'{u}blica 440, Santiago, Chile.}
\preprint{CECS-PHY-12/02, ESI-2372}

\begin{abstract}
Global charges and thermodynamic properties of three-dimensional higher spin
black holes that have been recently found in the literature are revisited.
Since these solutions possess a relaxed asymptotically AdS behavior, following
the canonical approach, it is shown that the global charges, and in particular
the energy, acquire explicit nontrivial contributions given by nonlinear terms
in the deviations with respect to the reference background. It is also found
that there are cases for which the first law of thermodynamics can be readily
worked out in the canonical ensemble, i.e., without work terms associated to
the presence of higher spin fields, and remarkably, the semiclassical higher
spin black hole entropy is exactly reproduced from Cardy formula.

\end{abstract}
\maketitle

%\pacs{}

\section{Introduction}

The exact higher spin black hole solutions in three spacetime dimensions that
have been recently found in \cite{GK, GKMP, CM} provide a unique arena in
order to acquire a deeper understanding of higher spin gravity
\cite{V1,V2,V3,V4}. These solutions possess a relaxed asymptotically AdS
behaviour as compared with the one proposed in refs. \cite{Henneaux-HS,
Theisen-HS, Campoleoni-HS, GH}. It is worth pointing out that a similar effect
occurs in the case of hairy black holes with scalar fields with slow fall-off
at infinity \cite{HMTZ-2+1,HMTZ-Log, HMTZ-D, HM, AM}. In this case, it is
known that the asymptotic conditions turn out to be relaxed with respect to
the ones of Brown and Henneaux \cite{Brown-Henneaux}, and as a consequence,
the global charges, and in particular the energy, acquire nontrivial
contributions given by nonlinear terms in the deviation of the fields with
respect to the reference background. Therefore, it is natural to wonder about
the persistence of this effect for black holes endowed with higher spin
fields. Here it is shown that this is indeed the case for the class of higher
spin black holes mentioned above. In fact, as they possess a relaxed
asymptotic behaviour, their energy does not depend linearly on the deviation
of the fields with respect to the background configuration. Therefore, the
higher spin black hole energy must be computed from scratch.

In the next section, higher spin gravity in three dimensions as a Chern-Simons
theory is revisited, where the canonical approach to construct conserved
charges as surface integrals is also briefly discussed. Section \ref{AGKP} is
devoted to perform the explicit computation of the energy, including the
analysis of nontrivial integrability conditions for the higher spin black hole
solution found in refs. \cite{GK,GKMP}. This is also carried out for the
solution of ref. \cite{CM} in section \ref{CMsect}, where it is also shown
that the first law of thermodynamics is automatically fulfilled in the
canonical ensemble, i.e., without work terms associated to the presence of
higher spin fields. The semiclassical entropy of this higher spin black hole
is also shown to be exactly reproduced by means of Cardy formula. Final
comments are discussed in section \ref{Discussion}.

\section{Higher spin gravity as a Chern-Simons theory in 3D}

As it was shown in \cite{Blencowe, BBS}, a Chern-Simons action whose gauge
group is given by $SL\left(  N,\mathbb{R}\right)  \times SL\left(
N,\mathbb{R}\right)  $ describes a theory of gravity with negative
cosmological constant, coupled to interacting fields of higher spin
$s=3,4,...,N$, in three dimensions. Analogously, a consistent theory of
gravity that includes the whole infinite tower of higher spin fields can be
constructed by means of two copies of the $hs(\lambda)$ algebra. Let us focus
on the simplest case that corresponds to $N=3$. The theory can then be
described in terms of two independent connection one-forms, $A^{+}$ and
$A^{-}$, associated to each copy of $SL\left(  3,\mathbb{R}\right)  $, so that
the action is given by%
\begin{equation}
I=I_{CS}\left[  A^{+}\right]  -I_{CS}\left[  A^{-}\right]  \ , \label{Itotal}%
\end{equation}
where%
\begin{equation}
I_{CS}\left[  A\right]  =\frac{k}{4\pi}\int_{M}\left\langle AdA+\frac{2}%
{3}A^{3}\right\rangle \ , \label{Ics}%
\end{equation}
and the level is determined by the Newton constant and the AdS radius
according to $k=\frac{l}{4G}$. Here the bracket stands for an invariant
nondegenerate bilinear form of $SL\left(  3,\mathbb{R}\right)  $ that is
proportional to the Cartan-Killing metric. The fundamental representation of
$SL\left(  3,\mathbb{R}\right)  $ is generated by $L_{i}$ and $W_{m}$, where
$i=-1,0,1$, and $m=-2,-1,...,+2$, and the bracket is given by a quarter of the
trace, i.e., $\left\langle \cdots\right\rangle =\frac{1}{4}\mathrm{tr}\left(
\cdots\right)  $, see e.g., \cite{Theisen-HS}.

Since the field equations imply the vanishing of $SL\left(  3,\mathbb{R}%
\right)  $ curvatures, i.e., $F^{\pm}=0$, the connections become locally flat
on shell.

It is useful to introduce a generalization of the dreibein and spin connection
according to%
\begin{equation}
A^{\pm}=\omega\pm\frac{e}{l}\ ,
\end{equation}
so that, in the principal embedding of $sl\left(  2,\mathbb{R}\right)  $ into
$sl\left(  3,\mathbb{R}\right)  $, the spacetime metric and the spin $3$ field
are recovered from%
\begin{equation}
g_{\mu\nu}=\frac{1}{2}\mathrm{tr}\left(  e_{\mu}e_{\nu}\right)  \ ,
\label{gmunu}%
\end{equation}
and%
\begin{equation}
\varphi_{\mu\nu\rho}=\frac{1}{3!}\mathrm{tr}\left(  e_{(\mu}e_{\nu}e_{\rho
)}\right)  \ , \label{phimunurho}%
\end{equation}
respectively.

It is worth pointing out that the metric transforms nontrivially under the
higher spin gauge symmetries embedded in $SL\left(  3,\mathbb{R}\right)
\times SL\left(  3,\mathbb{R}\right)  $, and as a consequence, some standard
geometric and physical notions may appear to be ambiguous, since they are no
longer gauge invariant. Therefore, in order to provide a reliable definition
of energy, our proposal is to stay attached to its very basic definition: the
energy corresponds to the conserved charge associated with the generator of
time evolution, i.e., it is given by the Hamiltonian.

\subsection{Canonical generators}

The suitable definition of energy we look for a Chern-Simons theory of the
form (\ref{Itotal}) can then be obtained following the Regge-Teitelboim
approach \cite{Regge-Teitelboim}. In the canonical formalism, the variation of
the conserved charge associated to an asymptotic gauge symmetry generated by
an algebra-valued parameter $\eta=\eta^{+}+\eta^{-}$ is given by%
\begin{equation}
\delta Q\left(  \eta\right)  =-\frac{k}{2\pi}\int_{\partial\Sigma}\left(
\left\langle \eta^{+}\delta A_{\phi}^{+}\right\rangle -\left\langle \eta
^{-}\delta A_{\phi}^{-}\right\rangle \right)  d\phi\ , \label{deltaqeta}%
\end{equation}
where $\partial\Sigma$ stands for the boundary of the spacelike section
$\Sigma\ $(see e.g., \cite{Balachandran, Banados-Q, Carlip-Q}). Since
diffeomorphisms are not independent of gauge transformations for a
Chern-Simons theory in three dimensions, the variation of the generator of an
asymptotic symmetry spanned by an asymptotic killing vector $\xi^{\mu}$,
reduces to%
\begin{equation}
\delta Q\left(  \xi\right)  =\frac{k}{2\pi}\int_{\partial\Sigma}\xi^{\mu
}\left(  \left\langle A_{\mu}^{+}\delta A_{\phi}^{+}\right\rangle
-\left\langle A_{\mu}^{-}\delta A_{\phi}^{-}\right\rangle \right)  d\phi\ .
\label{deltaqchi}%
\end{equation}
Therefore, the variation of the energy, $E=Q\left(  \partial_{t}\right)  $, is
given by%
\begin{equation}
\delta Q\left(  \partial_{t}\right)  =\frac{k}{2\pi}\int_{\partial\Sigma
}\left(  \left\langle A_{t}^{+}\delta A_{\phi}^{+}\right\rangle -\left\langle
A_{t}^{-}\delta A_{\phi}^{-}\right\rangle \right)  d\phi\ . \label{deltaM}%
\end{equation}
The variation of the canonical generators in eqs.\ (\ref{deltaqeta}),
(\ref{deltaqchi}), and (\ref{deltaM}) then corresponds to the ones of higher
spin gravity provided the parameter $\eta^{\pm}$ takes values on $hs\left(
\lambda\right)  $ or $sl\left(  N,\mathbb{R}\right)  $. In both cases, a
consistent set of asymptotic conditions has been proposed in refs.
\cite{Henneaux-HS} and \cite{Theisen-HS,Campoleoni-HS, GH}, respectively,
being such that the conserved charges turn out to be linear in the deviation
of the fields with respect to the AdS$_{3}$ background\footnote{The
supersymmetric extension of the asymptotic conditions of \cite{Henneaux-HS}
was developed in ref. \cite{Super-HS}.}. Indeed, for the case of $sl\left(
3,\mathbb{R}\right)  $, the asymptotic conditions for the gauge fields can be
written as%
\[
A^{\pm}=\bar{A}^{\pm}+\Delta A^{\pm}\ ,
\]
where the deviation with respect to the background configuration $\bar{A}%
^{\pm}$, which is assumed to be AdS$_{3}$ spacetime of radius $l$, is of the
form%
\begin{equation}
\Delta A^{\pm}=\pm\left(  -\frac{2\pi}{k}\mathcal{L}^{\pm}e^{-\rho}L_{\mp
1}^{\pm}\mp\frac{\pi}{2k}\mathcal{W}^{\pm}e^{-2\rho}W_{\mp2}^{\pm}\right)
dx^{\pm}\ , \label{DeltaA}%
\end{equation}
with $x^{\pm}=\frac{t}{l}\pm\phi$, and $L_{i}^{\pm}$ span two copies of the
$sl\left(  2,\mathbb{R}\right)  $ subalgebra. By virtue of (\ref{DeltaA}), it
is then simple to verify that the variation of the canonical generators in
(\ref{deltaqchi}) becomes linear in the deviation of the fields. Therefore,
the conserved charges can be readily integrated so that, in particular, the
zero modes of the Virasoro generators read%
\begin{equation}
L_{0}^{\pm}:=Q\left(  \partial_{\pm}\right)  =\int\mathcal{L}^{\pm}d\phi\ ,
\label{L0linear}%
\end{equation}
and hence, the energy (\ref{deltaM}) is given by%
\begin{equation}
E=Q\left(  \partial_{t}\right)  =\frac{1}{l}\int\left(  \mathcal{L}%
^{+}+\mathcal{L}^{-}\right)  d\phi\ . \label{MassLinear}%
\end{equation}
It is also simple to verify that eqs. (\ref{L0linear}) and (\ref{MassLinear})
also hold for the asymptotic conditions in \cite{Theisen-HS, Campoleoni-HS}
and \cite{Henneaux-HS, GH}, for $sl\left(  N,\mathbb{R}\right)  $ and
$hs\left(  \lambda\right)  $, where it has also been shown that the algebra of
the canonical generators acquires the same central extension as the one found
by Brown and Henneaux in \cite{Brown-Henneaux} for General Relativity with
negative cosmological constant,%
\begin{equation}
c=\frac{3l}{2G}\ . \label{C-BH}%
\end{equation}

In the next section we show that, since the higher spin black hole solutions
found in \cite{GK,GKMP,CM} do not fulfill the asymptotic conditions of ref.
\cite{Henneaux-HS, Theisen-HS}, their global charges, and then their energy,
differ from eqs. (\ref{L0linear}) and (\ref{MassLinear}), respectively,
because they acquire explicit nonlinear contributions that come from the
deviation of the fields with respect to the reference background. It is also
worth pointing out that further integrability conditions that guarantee that
the variation of the energy is an exact differential are also found. This
implies that some of the integration constants appearing on the higher spin
black hole solutions considered here generically become functionally related.

\section{Ammon-Gutperle-Kraus-Perlmutter solution}

\label{AGKP}

Let us first consider the higher spin black hole solution found in
refs.\ \cite{GK,GKMP} for the case $N=3$. The gauge field can be conveniently
written as%
\begin{equation}
A^{\pm}=g_{\pm}^{-1}a^{\pm}g_{\pm}+g_{\pm}^{-1}dg_{\pm}\ , \label{A+-}%
\end{equation}
where $g_{\pm}=g_{\pm}\left(  \rho\right)  $ stand for suitable elements of
each copy of $SL\left(  3,\mathbb{R}\right)  $ that depend only on the radial
coordinate, so that%
\begin{align}
a^{\pm}  &  =\pm\left(  L_{\pm1}^{\pm}-\frac{2\pi}{k}\mathcal{L}L_{\mp1}^{\pm
}\mp\frac{\pi}{2k}\mathcal{W}W_{\mp2}^{\pm}\right)  dx^{\pm}\nonumber\\
&  +\mu\left(  W_{\pm2}^{\pm}-\frac{4\pi}{k}\mathcal{L}W_{0}^{\pm}+\frac
{4\pi^{2}}{k^{2}}\mathcal{L}^{2}W_{\mp2}^{\pm}\pm\frac{4\pi}{k}\mathcal{W}%
L_{\mp1}^{\pm}\right)  dx^{\mp}\ ,
\end{align}
correspond to the connection in the \textquotedblleft wormhole
gauge\textquotedblright. As explained in \cite{GK,GKMP}, this solution does
not fulfill the asymptotic conditions of ref. \cite{Henneaux-HS, Theisen-HS}
in eq. (\ref{DeltaA}), since the metric asymptotically approaches to that of
AdS$_{3}$, but of radius $\tilde{l}=l/2$, and moreover the deviation with
respect to the background configuration, $\Delta A^{\pm}$, possesses
additional components along $dx^{\mp}=\frac{1}{\tilde{l}}dt\mp d\phi$.

It is simple to verify that, since $g_{\pm}=g_{\pm}\left(  \rho\right)  $, the
variation of the canonical generator in eq. (\ref{deltaM}) reduces to%
\begin{equation}
\delta Q\left(  \partial_{t}\right)  =\frac{k}{2\pi}\int_{\partial\Sigma
}\left(  \left\langle a_{t}^{+}\delta a_{\phi}^{+}\right\rangle -\left\langle
a_{t}^{-}\delta a_{\phi}^{-}\right\rangle \right)  d\phi\ , \label{DeltaQ-a}%
\end{equation}
and hence, the variation of the energy, $\delta E=\delta Q\left(  \partial
_{t}\right)  $, is given by%
\begin{equation}
\delta E=\frac{8\pi}{l}\left[  \delta\mathcal{L}-\frac{32\pi}{3k}%
\delta(\mathcal{L}^{2}\mu^{2})+\mu\delta\mathcal{W}+3\mathcal{W}\delta
\mu\right]  \ . \label{DeltaM-Kraus}%
\end{equation}

Since the configuration is static, by virtue of (\ref{deltaqchi}), one obtains
that%
\begin{equation}
\delta L_{0}^{\pm}=\delta Q\left(  \partial_{\pm}\right)  =\frac{\tilde{l}}%
{2}\delta E\ . \label{DeltaL0-Kraus}%
\end{equation}
According to eq. (\ref{DeltaM-Kraus}), the higher spin black hole energy not
only has the expected linear contribution in $\delta\mathcal{L}$, but also
acquires additional terms that depend nonlinearly in the integration
constants. Note that the energy is well defined provided its variation becomes
an exact differential, so that it can be integrated. Therefore, in order to
guarantee that, the last two terms at the r.h.s. of eq. (\ref{DeltaM-Kraus})
give a nontrivial integrability condition that have to be fulfilled. This
condition then reads%
\begin{equation}
\delta^{2}E=\frac{16\pi}{l}\ \delta\mathcal{W}\wedge\delta\mu=0\ ,
\end{equation}
and as a consequence, the integration constants $\mu$ and $\mathcal{W}$ are
not generically independent, i.e., they become functionally related as%
\begin{equation}
\mu=\mathcal{F}^{\prime}\left(  \mathcal{W}\right)  \ , \label{mu(W)}%
\end{equation}
where $\mathcal{F}^{\prime}$ is the derivative of an arbitrary function
$\mathcal{F}$ that is fixed once precise boundary conditions are provided.
Hence, the energy is given by
\begin{equation}
E=\frac{8\pi}{l}\left[  \mathcal{L}-\frac{32\pi}{3k}\mu^{2}\mathcal{L}%
^{2}+3\mu\mathcal{W}-2\mathcal{F}\right]  \ , \label{Mass-Kraus}%
\end{equation}
up to an arbitrary constant without variation. Note that in the case of
$\mu=\mu_{0}$, where $\mu_{0}$ is an arbitrary constant without variation,
which corresponds to the one considered in \cite{GK,GKMP}, the function
$\mathcal{F}$ becomes fixed according to $\mathcal{F}=\mu_{0}\mathcal{W}$, so
that the energy reduces to%
\begin{equation}
E=\frac{8\pi}{l}\left[  \mathcal{L}-\frac{32\pi}{3k}\mu_{0}^{2}\mathcal{L}%
^{2}+\mu_{0}\mathcal{W}\right]  \ . \label{Energy-mu0}%
\end{equation}

It is worth pointing out that the explicit dependence of the energy in terms
of the function $\mathcal{F}$, which is precisely specified through the fixed
data at the boundary, e.g., through Dirichlet, Neumann, mixed or generic
nonlinear boundary conditions, is an effect that is known to occur when the
fields possess a relaxed asymptotic behaviour, being such that the global
charges acquire nontrivial contributions given by nonlinear terms in the
deviation of the fields with respect to the background configuration. Indeed,
this is the generic case for scalar fields with relaxed AdS asymptotics, which
has been widely discussed in Refs. \cite{HMTZ-2+1,HMTZ-Log, HMTZ-D, HM,
AM,DG0,DG1,DG2,DG3,DG4,DG5}. Further details concerning with the precise
fixing of the function $\mathcal{F}$ are discussed in section \ref{Discussion}.

\section{Castro-Hijano-Lepage-Jutier-Maloney solution}

\label{CMsect}

The second example we consider is the static higher spin black hole solution
found in \cite{CM}\footnote{As pointed out to us by an anonymous referee, this
solution should not be named \textquotedblleft higher spin black
hole\textquotedblright, in the sense that it carries a $U(1)$ spin 1 charge
instead of a higher spin charge.}, also for $N=3$. As in the previous case, it
is convenient to express the gauge field as in eq. (\ref{A+-}), where now
$g_{\pm}=e^{\pm\rho L_{0}^{\pm}}$, and%
\begin{equation}
a^{\pm}=\pm\left(  \ell_{P}L_{\pm1}^{\pm}-\mathcal{L}L_{\mp1}^{\pm}\pm\Phi
W_{0}^{\pm}\right)  dx^{\pm}+\left(  \ell_{D}W_{\pm2}^{\pm}+\mathcal{W}%
W_{\mp2}^{\pm}-QW_{0}^{\pm}\right)  dx^{\mp}\ ,
\end{equation}
with\footnote{We have chosen a different orientability as compared with the
one in \cite{CM}, i.e., $x^{+}\leftrightarrow x^{-}$. As in the previous
section, here $x^{\pm}=\frac{t}{\tilde{l}}\pm\phi$, and hereafter we will
consider the branch with $\ell_{P}$,$\ \ell_{D}\neq0$.}%
\begin{equation}
Q\ell_{P}-2\mathcal{L}\ell_{D}=0\ \ ;\ \ Q\mathcal{L}-2\mathcal{W}\ell
_{P}=0\ .
\end{equation}

It is simple to verify that this solution does not fit within the asymptotic
conditions in eq. (\ref{DeltaA}). Indeed, the asymptotic form of the metric
approaches to AdS$_{3}$ of radius $\tilde{l}=l/2$, and the deviation of the
gauge field $A^{\pm}$ also possesses additional components along $dx^{\mp}$.

The variation of the energy can then be obtained from eq. (\ref{DeltaQ-a}),
which reads%
\begin{equation}
\delta E=\delta Q\left(  \partial_{t}\right)  =\frac{1}{3G}\left[
\delta\left(  3\mathcal{L}\ell_{P}-4Q^{2}-2Q\Phi+\Phi^{2}\right)  +4\Phi\delta
Q\right]  \ . \label{DeltaM-Castro}%
\end{equation}
Therefore, as expected, the variation of the energy has a nonlinear dependence
in the integration constants. The integrability condition that comes from
(\ref{DeltaM-Castro}) is given by%
\begin{equation}
\delta^{2}E=\frac{4}{3G}\delta\Phi\wedge\delta Q=0\ ,
\end{equation}
which means that the integration constants $\Phi$ and $Q$ are functionally
related. It is convenient to express this relation as%
\begin{equation}
\Phi=\mathcal{F}^{\prime}(Q)\ , \label{F(Phi)}%
\end{equation}
for some arbitrary function $\mathcal{F}$. Hence, up to an arbitrary constant
without variation, the energy can be written as%
\begin{equation}
E=\frac{1}{G}\left[  \mathcal{L}\ell_{P}+4Q^{2}\right]  +w(Q)\ ,
\label{MassCM}%
\end{equation}
with%
\begin{equation}
w(Q):=\frac{1}{3G}\left[  (\Phi-4Q)(\Phi+4Q)-2(Q\Phi-2\mathcal{F})\right]  \ ,
\label{w}%
\end{equation}
so that%
\begin{equation}
L_{0}:=L_{0}^{\pm}=Q\left(  \partial_{\pm}\right)  =\frac{\tilde{l}}{2}%
E=\frac{l}{4}E\ . \label{Lo-CM}%
\end{equation}

\subsection{Thermodynamics}

As it was shown in \cite{CM}, requiring the holonomy around the thermal cycle
of the Euclidean solution to be trivial, allows to fix Hawking temperature
according to%
\begin{equation}
T=\frac{2}{\pi l}\sqrt{\mathcal{L}\ell_{p}+4Q^{2}}\ , \label{T-CMq}%
\end{equation}
and gives an additional condition that reads
\begin{equation}
\Phi=4Q\ . \label{phi4q}%
\end{equation}
This restricts the precise form of the function $\mathcal{F}$ in
(\ref{F(Phi)}), according to $\mathcal{F}=2Q^{2}$. As a consequence, the
function $w(Q)$ in (\ref{w}) vanishes, so that the energy becomes proportional
to the square of the temperature, i.e.,%
\begin{equation}
E=\frac{\pi^{2}l^{2}}{4G}T^{2}\ , \label{M(T)}%
\end{equation}
and then the entropy can be readily found in the canonical ensemble, $dE=TdS$,
to be given by%
\begin{equation}
S=\pi l\sqrt{\frac{E}{G}}\ . \label{S(M)}%
\end{equation}
Note that, according to eq. (\ref{Lo-CM}), the semiclassical entropy of the
higher spin black hole (\ref{S(M)}) reads%
\begin{equation}
S=4\pi\sqrt{\frac{l}{4G}L_{0}}\ , \label{S(L0)}%
\end{equation}
which exactly agrees with Cardy formula,%
\begin{equation}
S=4\pi\sqrt{\frac{c}{6}L_{0}}\ , \label{Cardy standard}%
\end{equation}
provided
\begin{equation}
c=\frac{3l}{2G}\ ,
\end{equation}
i.e., precisely the standard central charge that has been found to hold also
for higher spin gravity with asymptotically AdS$_{3}$ boundary conditions
\cite{Henneaux-HS, Theisen-HS, Campoleoni-HS, GH}. This result suggests that
the higher spin black hole found in \cite{CM} could be naturally regarded as a
large nonperturbative deviation with respect to the AdS$_{3}$ vacuum of radius
$l$.

\section{Discussion}

\label{Discussion}

The canonical formalism to compute conserved charges as surface integrals
\cite{Regge-Teitelboim} has been briefly reviewed in the case of higher spin
gravity in three dimensions, and it was applied in order to obtain the energy
of the higher spin black holes in refs. \cite{GK,GKMP} and \cite{CM}. In both
cases, it was found that the energy acquires nonlinear terms in the deviation
of the fields with respect to the reference background. This goes by hand with
non trivial functional relationships between the integration constants that
have to be fulfilled in order to ensure integrability of the charges. This
effect occurs due to the fact that these solutions possess a relaxed
asymptotically AdS behavior as compared with the ones in \cite{Henneaux-HS,
Theisen-HS, Campoleoni-HS, GH}, and hence the global charges, in particular
their energy, differ from eqs. (\ref{L0linear}) and (\ref{MassLinear}%
)\footnote{Further examples exhibiting similar features have also been
discussed in refs. \cite{HMTmore, HMTwarped} in the context of topologically
massive gravity \cite{Deser:1982vy,DeserJT,DeserCosmo}, as well as in
\cite{PTT} for BHT \textquotedblleft new\textquotedblright\ massive gravity
\cite{BHT} in vacuum.}. It is worth then pointing out that finite charges as
surface integrals that are obtained through different perturbative approaches
do not capture this effect. Indeed, although they may transform suitably under
the Virasoro symmetry, \textit{a priori}, there is no guarantee that they
reproduce the energy unless one explicitly check that they generate the time evolution.

For the specific examples considered here, some remarks are in order. In the
case of the higher spin black hole of \cite{GK,GKMP} the integrability
condition that makes the variation of the energy to be an exact differential
reduces to eq. (\ref{mu(W)}). The case $\mu=\mu_{0}$, where $\mu_{0}$ is
arbitrary and without variation is certainly one of the possibilities, and
hence along the lines of the AdS-CFT correspondence, the generic functional
relationship (\ref{mu(W)}) naturally arises in the context of multi-trace
deformations \cite{KW,W}. It is also worth mentioning that, as explained in
\cite{GK,GKMP}, for the Euclidean solution, requiring the holonomy around the
thermal cycle to be trivial not only determines the higher spin black hole
temperature, but also gives an additional condition. As a consequence, once
eq. (\ref{mu(W)}) is imposed, the energy in eq. (\ref{Mass-Kraus}) turns out
to depend on a single integration constant. Further subtleties concerning how
thermodynamics works once these results are taken into account for this and
other examples are discussed in \cite{PTT-Entropy}.

In the case of the higher spin black hole found in \cite{CM}, requiring
triviality of the holonomy around the thermal cycle singles out a unique
possibility for the functional relationship in eq. (\ref{F(Phi)}). Since the
energy becomes proportional to the square of the temperature, as in eq.
(\ref{M(T)}), the thermodynamics can be readily carried out in the canonical
ensemble. Remarkably, the semiclassical entropy of the higher spin black hole
was shown to exactly agree with Cardy formula, provided the central charge is
the one of Brown and Henneaux, which was found to hold also for higher spin
gravity with asymptotically AdS$_{3}$ boundary conditions \cite{Henneaux-HS,
Theisen-HS, Campoleoni-HS, GH}. Note that if the asymptotic conditions could
be relaxed in a consistent way with the asymptotic symmetries, the central
charge would not change, since it is determined by the the ground state
configuration. Therefore, although the higher spin black hole of \cite{CM}
asymptotically approaches to AdS$_{3}$ spacetime of radius $\tilde{l}=\frac
{l}{2}$, our result naturally suggests that it could be consistently regarded
as a large nonperturbative deviation with respect to the AdS$_{3}$ vacuum of
radius $l$. In this sense, it is worth mentioning that AdS$_{3}$ with radius
$l$ has been argued to be the only ground state for which the perturbative
spectrum of the theory could admit a unitary representation for large values
of the central charge \cite{CH-Unitarity}.

Curiously, in this sense, the usual practice that suggests the possibility of
regarding the asymptotics of the higher spin black hole as a perturbative
deviation with respect to AdS$_{3}$ of radius $\tilde{l}$ does not appear to
be an appealing one. Indeed, if this was the case, according to ref.
\cite{GKMP}, the central charge that corresponds to the ground state would be
given by $\tilde{c}=\frac{3l}{8G}$, and hence the standard form of Cardy
formula in eq. (\ref{Cardy standard}), with $c=\tilde{c}$, would fail in
reproducing its semiclassical entropy. It is also worth pointing out that in
this case, along the lines of refs. \cite{CMT1, CMT2}, Cardy formula still
would have a chance to work if it is generalized so as to admit nontrivial
lowest eigenvalues for the Virasoro operators, which would correspond to the
ones of a suitable ground state that asymptotically approaches to AdS$_{3}$ of
radius $\tilde{l}$. However, in this case, the lowest eigenvalues of the
Virasoro operators would be given by $\bar{L}_{0}^{\pm}=-\frac{l}{16G}%
<-\frac{\tilde{c}}{24}$, which manifestly violates the unitarity bound.

As an ending remark, it would be interesting to explore the possibility of
recovering the results discussed here from the dual theory at the boundary,
along the lines of \cite{H1,H2,H3,H4,H5,H6}, as well as inspecting whether
they persist for the different class of higher spin black holes found in
\cite{PK, TAN, GGR}, or for the generic solution of \cite{Banados-Theisen}.

\acknowledgments We thank H. Afshar, G. Barnich, X. Bekaert, E. Bergshoeff, N.
Boulanger, C. Bunster, A. Campoleoni, A. Castro, M. Gary, H. Gonz\'{a}lez, M.
Grigoriev, D. Grumiller, O. Hohm, C. Iazeolla, P. Kraus, S. Pfenninger, R.
Rahman, R. Rashkov, P. Sundell, M. Valenzuela, A. Waldron, J. Zanelli, and
specially to M. Henneaux, C. Mart\'{\i}nez and S. Theisen for many useful
discussions and comments. We also wish to thank the organizers of the Workshop
on Higher Spin Gravity, hosted by the Erwin Schr\"{o}dinger Institute (ESI),
during April 2012 in Vienna, for the opportunity of presenting this work. We
thank the kind hospitality at the Physique th\'{e}orique et math\'{e}matique
group of the Universit\'{e} Libre de Bruxelles, the International Solvay
Institutes, and the Max-Planck-Institut f\"{u}r Gravitationsphysik
(Albert-Einstein-Institut). This work has been partially funded by the
Fondecyt grants N${^{\circ}}$ 1130658, 1121031, 3110122, 3110141. The Centro
de Estudios Cient\'{\i}ficos (CECs) is funded by the Chilean Government
through the Centers of Excellence Base Financing Program of Conicyt.

%\appendix

\end{document}